# Discomfort Monitoring System for Residential Electrical Water Heater

Ziyad Almajali

*PhD, Department of electrical engineering, Mutah university, Karak, Jordan*

ziyad@mutah.edu.jo

***Abstract*** — *An approach is described in this work for detecting discomfort moments during electrical water heater daily usage. The approach employs chromatic analyzing sensors signals of electrical water heater systems for producing distinguishable mapping to characterize and identify selected discomfort moments. The preliminary results obtained indicate that it is possible to distinguish such events in a non-intrusive approach.*

*Hot water comfort detection and classification intelligently through human behavior monitoring and analyses plays an important role in both, energy saving and energy management.*

*The focus is on recording the discomfort situations followed by merging system outcomes with efficiency evaluation will be used to provide helpful recommendations for selecting appropriate operating strategy.*

## I. INTRODUCTION

A high percentage of electrical energy is consumed in the domestic sector and a large share of this consumption is due to the usage of the electrical water heater. Statistics in Jordan show that 40% of the consumed energy is used in homes. Homeowners prefer to use electrical water heaters over traditional methods for suppling their daily hot water requirements in 56% of the houses [1]. They select water heaters for comfort as its fast and clean energy source, but its usage results in the large portion of the sector monthly paid utility bills.

The primary purpose of using electrical energy is to serve the user, but there is an ongoing search for energy saving opportunities, especially with appliances that account for the largest share of the bill, such as water heaters [2]. Although, following such policies are among the main goals of future smart homes energy management systems, but this may threaten the comfort level of homeowners.

A solution that combines energy savings without compromising user comfort can be planned by relying on smart appliances; that are the main part of the smart home vision, which is now an active area of interest for researchers. Nowadays, electrical appliance industries are shifting to fully smart device production in preparation to join various smart grid technologies such as smart home and demand response. But it is noticeable that this trend has become active in electronic appliances with low energy consumption, while priority should be given to appliances with high energy consumption, such as electric water heaters.

The daily operation of the heater falls within the two options; the first option is to keep the device in a state of continuous operation. The second option is to turn it on when and only when hot water is needed. To answer the question of which operating strategy is preferred for producing reduced utility bill, homeowners' answers are already divided between the two popular operating strategies. Some users claim that it is economically preferred to leave the power ON for all the time, and other users claim that a lower bill will result from turning the power ON only when hot water is required. The described tool in this work may help in answering such question as detected level of comfort/discomfort when combined with information about system efficiency, may lead to an advice of continuing with the current operating strategy or advice of immediate change to another option.

Generally, for each of the traditional adopted operating control strategies there are advantages and there are disadvantages. From an economic point of view homeowners are divided into two groups in deciding which strategy is better than the other. Both groups got no scientific and accurate proof on their claims because it is simply based on comparison between bills of two consecutive months. The comparison is not accurate because each bill for each month is affected greatly by that month's usage and the weather conditions in that month. Moreover, the monthly bill is recording the consumed power for all the house appliances together.

Despite the desire to keep the user comfortable, the path to this should not lead to their inconvenience. thus, it is necessary to get accurate interpretation of human action and behavior through a tool that can perform intelligent comfort level detection [3]. Merging such tool with system efficiency performance information will provide useful and important usage advice which will extremely help the decision-making stage tools to better understand the situation and thus selecting appropriate decisions which will be reflected in comfortable usage in addition to optimum savings.

Comfort Detection is a difficult object to achieve. Most of the research in this field does not focus on water heaters, but rather revolves around HVAC systems and trying to discover the discomfort of their users in a non-intrusively. [4-6] and most of them depend on analysing signals from his face or chest or even by measuring his body temperature using thermal cameras. Most of these methods focus on human comfort, rather, than energy savings

For comfort factor indication, some methods may be based on direct confrontation such as multiple direct questionnaires to estimate user comfort which is found annoying and instead of comfort level estimation, such questionnaire will result in decreasing comfort due to continues and multiple survey filling requests. [7]

For energy savings and best operation strategy selection, most of the developed strategies and solutions for electrical water heaters, such as rule based [8] and the fuzzy logic [9] [10] are based and affected by the adopted and implemented pricing tariffs. Peak shifting strategies were adopted in [11] [12][13]. In most of these works, the human factor was not considered, and the focus was on the savings rather, than the comfort.

In this paper, the proposed approach aims for human behavior recognition intelligently through chromatic analyzing of water heater system signals sensors like water flow, electrical input power, thermal power in addition to the manual switching on and switching off actions by system owner. The result will be combined with heater efficiency information, thus the decision will be based on both factors: the human comfort and energy savings as well.

Few categorized situations will be detected and distinguished from the general behavior of the user's daily life. The technique algorithm outcome may then be input to a simple classification stage using approaches such as fuzzification or even simple program algorithm.

## II. MONITORED SYSTEM

Toward smart monitoring system for a residential electrical water heater, a complete set of sensors and meters should be employed to monitor, record, and log all its vital signals.

Starting from the electrical energy entering the device. It is advisable to install a special power meter, figure 1 shows a sample of heater absorbed power log for a duration of thirty-six minutes. The total required energy for this operation is evaluated by calculating the total area under the curve with the unit of KWh.

Figure 1 indicates the monitored water heater absorption slight fluctuation around 453 watts with ±10 watts variation, which is due to the system voltage variation with other load changes, but due to its low value, it can be ignored.

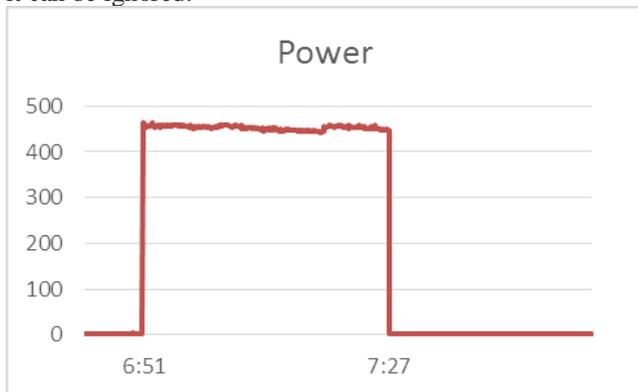

**Fig 1: sample of heater-absorbed power log**

A data logger is required for data collection with its time stamp. Another lower cost solution is using a cheap clamp meter instead of the costly wattmeter with an assumption of constant voltage at the device terminals. The accuracy of the reading in this method is questionable due to the incorrect assumption of voltage constancy. As it is noticeable, as shown in figure 2, that the voltage suffers from a clear fluctuation. It may fluctuate within the permissible limits, but the fluctuation is clear, especially at the moment (a), moment of device operating, which leads to a clear decrease in the voltage, and the moment (b), moment of turning off, which clearly leads to an increase of the voltage. All this leads to preferring the wattmeter method over the option of installing a clamp ammeter. Another option is using two meters, one for the current and the second one for the voltage, and the power calculation will be based on the ohmic type of heater load by direct multiplication of the two readings.

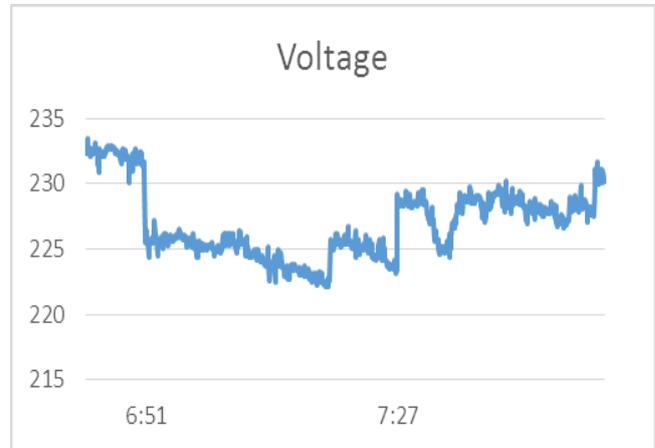

**Fig 2: sample of voltage log at heater terminals**

In addition to the electrical energy, the temperature of the water which is entering the heater may be of significant value as it will hold information about how much energy will be required to supply it with to reach the heater set point temperature. Cold water will need much more energy than that with normal room temperature. Water temperatures depend on the season and weather conditions.

Measurement of how much water is entering the system is also required, but since it is the same amount that is leaving the system, then a single flow meter at the outlet tap of the hot water will be enough. And a temperature sensor will be required for estimating the output power. The cold-water tap is not part of the system, but measuring its flow amount will be a required parameter for detecting some discomfort moments; thus, another flow meter at the cold-water tap may be required. Figure 3 shows a sample of water flow log for a duration of eight minutes.

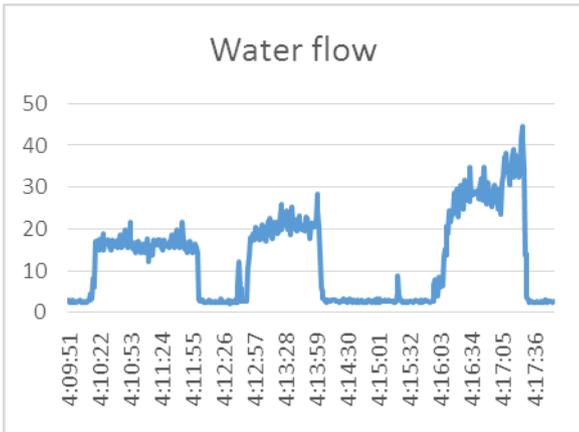

**Fig 3: sample of water flow log**

## III. COMFORT INDICATOR DETECTING SYSTEM

Occurrence of moments of discomfort when using the electrical water heater are probable and could occur frequently throughout the user energy consumption profiles. If a questionnaire is used as the day end for the purpose of providing the complete record of these events for future usage, the user may not remember all these moments, as some events may slip from his memories. Multiple questionnaires as well as the Immediate recording requirement may worsen the case as the user may consider both options annoying and considering them as another source of discomfort to his daily life.

The adopted solution in this work is by monitoring such events automatically with no direct intervention. The monitoring should be designed and aim eventually for generating of energy management advice, regard the electrical water heater operating conditions; this advice will be based on the fusion of the detected level of comfort together with the information of efficiency calculations tool.

A few selected discomfort/comfort targeted situations will be illustrated in subsection A. The recognition system consists of three parts; the first part is chromatic transformation of sensor signals into the xyz chromatic map. The chromaticity is introduced in subsection B. While the second part is regarding the used chromatic mapping and namely the xyz mapping algorithm which will be introduced in section C. the third part will describe the proposed system algorithm and implementation and this will be in subsection D with samples of the obtained classification results by chromatically processing of the captured signals.

### A. Discomfort situations

It is hard to extract and analyze detailed behavioral characteristics of comfort and/or discomfort moments for the user with the only provided data and the available input information. but a simple behavior monitor for specific events could be built which will focus only on the moments of discomfort.

This monitor tool will depend on how most of homeowners would deal with such events. usually which they could face and deal with it without sometimes even announcing how annoyed they are. three examples of selected discomfort situations are illustrated in the following subsections.

*a) Case 1*

Moments of discomfort sometimes appear when the user requests hot water. The current situation occurs when user opens the hot water tap seeking for the hot water but find the coming out water cold. Usually, the annoyed user will directly turn the heater on. Figure 4 illustrates the mentioned situation with the described events sequence. This case is a discomfort moment because the user did not get the hot water when he needed it and turning the heater on is proving the discomfort situation occurrence. The heater, of course, will need time to raise the water temperature to the required level that suits the user. And the waiting is unpleasant for any user.

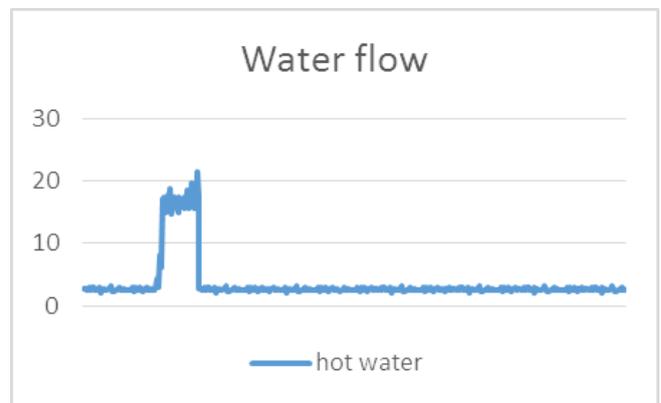
(a)

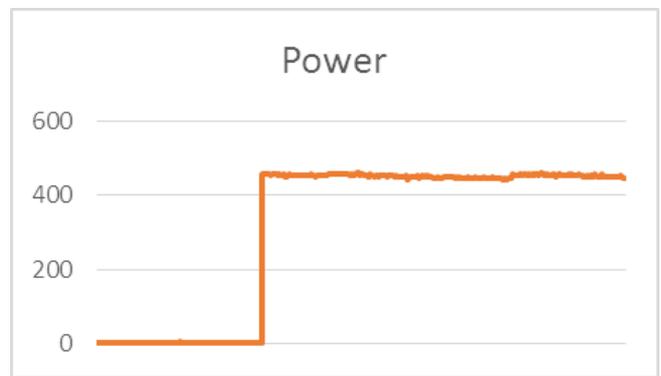
(b)
**Fig 4: Case 1**

It is worth noting that the situation mentioned here usually occurs in the second type of operation mentioned in the introduction; i.e. (to turn the heater on when and only when hot water is needed). Recording such events frequently may mean that the user is not happy with this option of operating and not feeling comfortable.

Figure 4.a shows the hot water opening moment and figure 4.b records the heater turning on moment through absorbed power monitoring.

And it can be recorded as a comfort situation when monitored signals indicate using a good amount of hot water without turning the heater on. Figure 5 shows an example of such a situation.

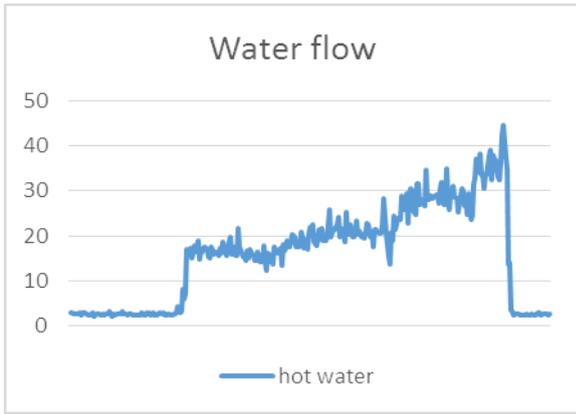

(a)

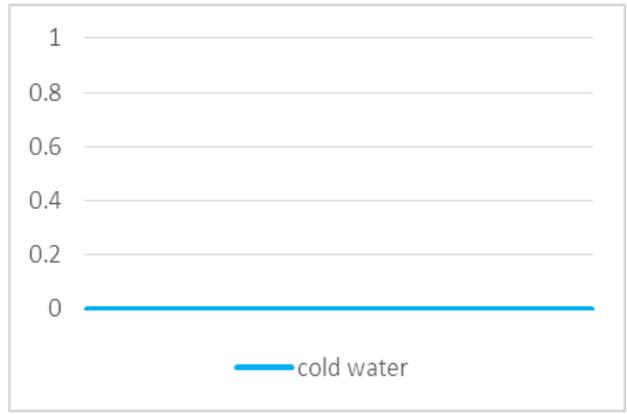

(b)

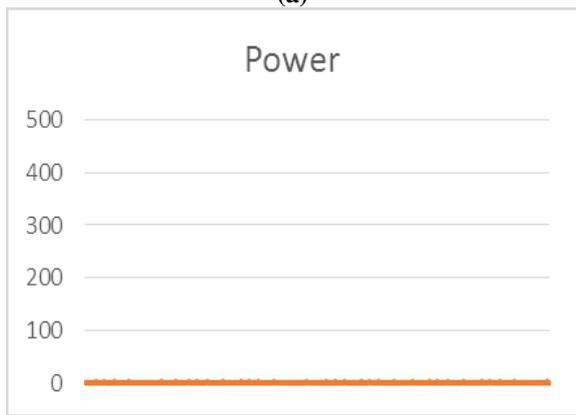

(b)
**Fig 5: comfort situation example**

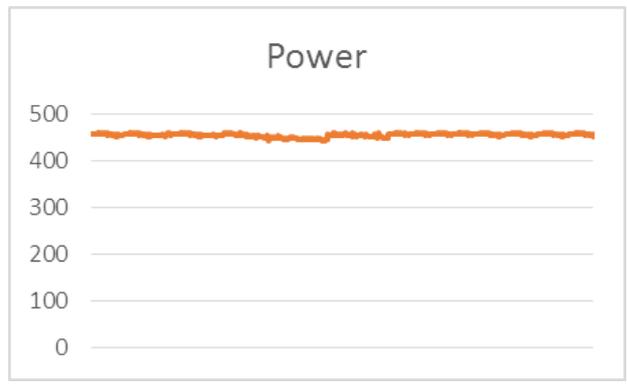

(c)
**Fig 6: case 2**

*b) Case 2*

Another discomfort Moment is related to hot water requests. Which may be noticed when user open the hot water tap fully, although the heater still in the heating stage and the temperature did not reach the set point, yet. what may prove the discomfort Moments occurrence is the fully opening hot water tap without opening the cold water tap as illustrated in figure 6. Detection of such events requires monitoring and analyzing signals from hot water taps, cold water taps and water temperature.

Figure 6.a shows the hot water continuous absorption; figure 6.b shows how no cold water was required. and figure 6.c shows the low temperature of the absorbed water. The last figure may be replaced by monitoring input power absorption, which will serve the same purpose.

*c) Case 3*

An example of a discomfort event is finding the water too hot and beyond the user's desirable heat level. The user may open the two water taps together, cold, and hot, at the same time. Although this case means the user's desire for a heavy flow of water, it may also mean that the user is not comfortable with the high temperature and opening the cold-water tap means the desire to adjust the temperature. To confirm the state of discomfort, the amount of flow from the two taps should be monitored, which is proven if the system detects more cold water than hot water, as illustrated in figure 7.

When this condition recurs, its treatment comes by adjusting the set temperature in the heater to a lower degree.

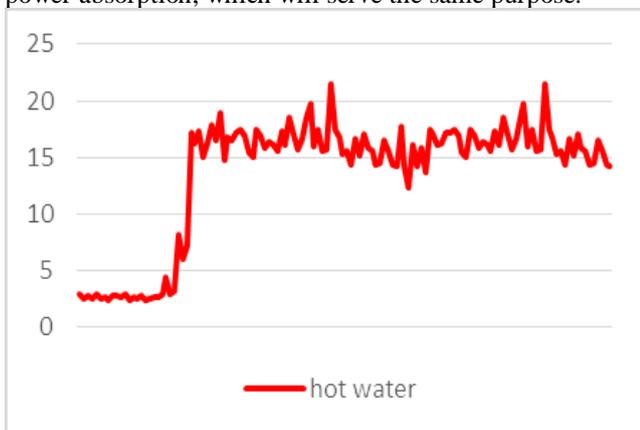

(a)

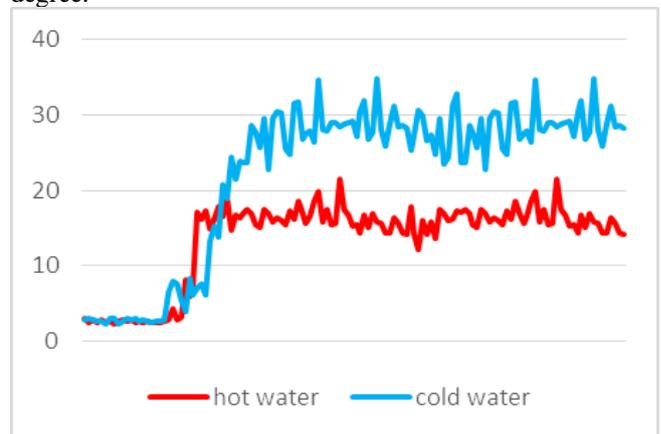

**Fig 7: case 3**

## B. Chromatic Principles

Complex events evolvement monitoring by the eye are detected and classified by the brain processing stage easily. color vision by the eye is the original source of chromaticity principle.

To explain the chromatic idea. Figure 8 shows two signals which are covered by three signal processors (R, G and B), whose responses are overlapping to cover the entire signals range. the same idea is taken from the eye different color filters covering the scene in front of it and sending the received signal to the brain processing stage. The transformation and processing of the processor's outcomes enhances the possibility of detecting an emerging event.

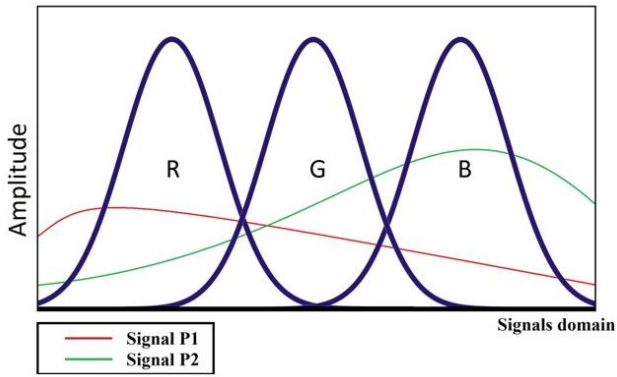

**Fig 8:**

The three signal processors outcomes are given by:

$$R = \int R(\phi)P(\phi)\,d\phi \quad (1)$$
$$G = \int G(\phi)P(\phi)\,d\phi \quad (2)$$
$$B = \int B(\phi)P(\phi)\,d\phi \quad (3)$$

$P(\phi)$ is the monitored signal and $R(\phi), G(\phi)$ and $B(\phi)$ are the processors profiles. $\phi$ is the monitored signals domain.

## C. xyz chromatic transformation

The signal related to a monitored event needs to be covered entirely by color processors; the outputs are then processed by appropriate transformations. there are various variations of this chromaticity transformations such as the xyz which is adopted in this work. The original transformation algorithm is shown in equations 4-7 where relative magnitudes (x, y, z) for each of the three overlapping processor outputs (R, G, B) in addition to the signal strength index (L) can be evaluated.

$$x = \frac{R}{3L} \quad (4)$$

$$y = \frac{G}{3L} \quad (5)$$

$$z = \frac{B}{3L} \quad (6)$$

$$L = \frac{R + G + B}{3} \quad (7)$$

information from the monitored event can be reflected in the relative magnitudes (x, y, and z) and the equation shows the relation between the different processors in producing of them. a graphical representation of two of the relative magnitude could hold enough information for the classification purpose, thus a two-dimensional figure will be selected instead of three dimensional one. For the RGB processor, triangular form has been chosen for its responses for simplicity, and a relation between x and y will be selected from the xyz since it was found adequate for identifying significant differences from which distinguishing features may be indicated.

The original algorithm can be modified according to the application as long as modification helps in accurate classifications and depending on the monitored events.

## D. proposed system

To explain the operating strategy effect on monitoring triggering, let us take an example. In case 2, and if working in the continuous operation mood, the device will be already turned on and the trigger for the event is hot water tap opening. the significant factor for discomfort detection is the temperature of the absorbed water. While, in case 1, and working in the (turn the heater when required) operation mood, i.e., the second mood, the device is turned on when hot water is required only. Thus, the trigger here is for the event of turning it on after a short hot water tap opening, and the significant factor here is the timing sequence of the two events as illustrated in case 1.

The proposed system is built on monitoring sensor signals chromatically. R, G and B filters profile are imposed over, and the outcome are fed to the chromatic processing stage.

The system signals are the amount of hot water, the amount of cold water, and the electrical power drawn. Due to the specificity of the monitored system and the multiplicity of received signals, a preprocessing stage is applied to the input signals. As shown in the flowchart in figure 9. a slight modification was also necessary to be applied to the original algorithm of the x,y and z parameters to suit the monitored conditions. The modified algorithm equations are given in equations 9-11

$$x = \frac{R}{abs(3L)} \quad (9)$$

$$y = \frac{G}{abs(3L)} \quad (10)$$

$$z = \frac{B}{abs(3L)} \quad (11)$$

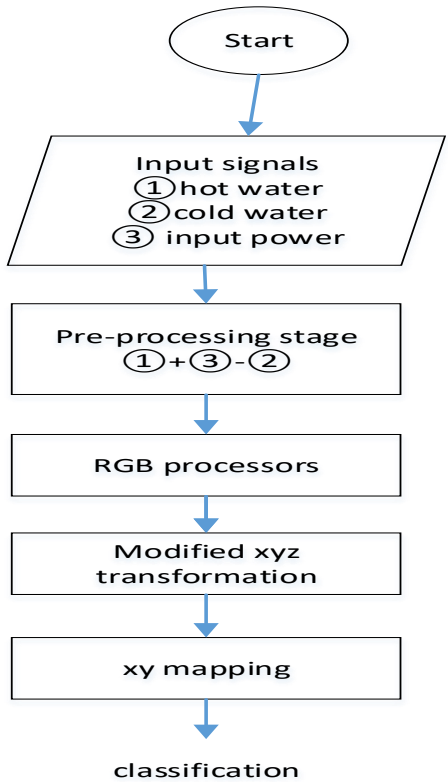

**Fig 9: method flowchart**

When the algorithm is applied, three variables are created, x, y, and z. In view of the interrelationship between the variables, which is clear in the formulation of their equations. It was sufficient to draw the relationship between only two variables namely x and L to express the clear difference between the three monitored cases, as shown in figure 10.

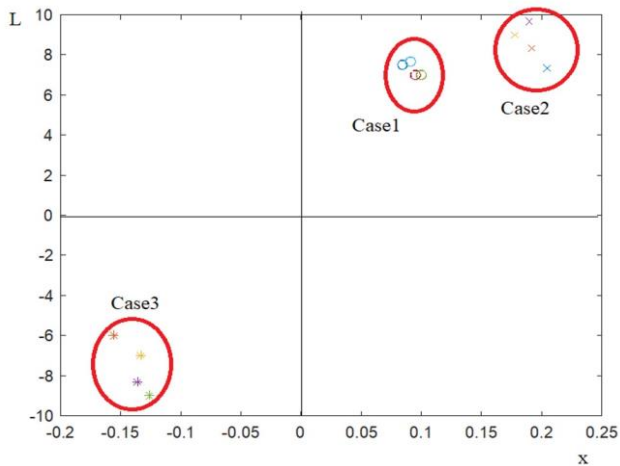

**Fig 10: x-L mapping diagram**

## IV. DISCUSSION

Repeatability between test cases with different usage levels and with random slightly different timing between events leads to the formation of clusters which are distinguishable. Samples of these results are shown on the chromatic x-L mapping diagram in figure 10, which shows how different monitored situations are distinguishable from each other with sufficient robustness.

The object of the designed tool is not counting of the discomfort situations, but rather to distinguish between them, given the difference between heaters characteristics from each other and the different nature of usage behavior among homeowners and to ensure a high-accuracy that has the ability to distinguish different cases regardless of the type of heater and regardless nature of use It is recommended that the classification be addressed in more depth through another classification system such as a fuzzifier system or a neural network-based system, and this will be deferred to future work.

In another work related to the same current project [14], the researcher created a tool to show the efficiency of the heater and monitor it continuously. Great benefits may arise from linking the two research outputs together.

The situation can be described as ideal or optimum if high efficiency is obtained from a system with few records of moments of discomfort. Here there is no need to change any thing in the system. But if the moments of discomfort are few, and the efficiency of the system is low. It is assumed here that the losses are significant due to poor insulation, especially in the case of continuous operating mood, as it is left without withdrawal and for long periods. Here, the user is advised to switch the operating mood and try to operate the system when needed for losses reduction, and it is also possible to advise for a heater change.

If significant moments of discomfort are recorded and this coincides with the recording of low system efficiency, the situation here requires urgent action such as a complete device replacement. While if significant moments of discomfort are recorded but the system efficiency is found satisfying, here, the advice will be regarding the operation mood switching with the hope of reducing moments of discomfort.

## V. CONCLUSIONS

Many users prefer electric water heaters over other types because they want to have the convenience of use, but it is reflected in a higher bill which forces some of them to modify the method of operation in order to save at the expense of the comfort.

A non-intrusive approach is described in this work for detecting discomfort moments during electrical water heater daily usage. The approach employs chromatic analysis of sensors signals of electrical water heater systems for producing distinguishable mapping to characterize and identify selected discomfort moments.

The focus is on recording the discomfort situations followed by merging system outcomes of efficiency evaluation will be used to provide helpful recommendations for selecting appropriate operating strategy that will serve both good savings and comfort usage.

## ACKNOWLEDGMENT
This research was a fully funded project by Mutah University research support deanship, titled "Optimal operating control strategy for domestic electrical water heaters"